\documentclass[10pt,pra,amsmath,amssymb,aps,showkeys,showpacs,twoside,final,secnumarabic,nofootinbib]{revtex4-2}
\usepackage{cmap,xcolor} 
\usepackage[utf8]{inputenc}
\usepackage[unicode=true,colorlinks=true,linkcolor=blue, urlcolor=teal, citecolor = blue,breaklinks]{hyperref}
\usepackage{graphicx}
\usepackage{bm}
\usepackage{multirow}
\usepackage{url}
\DeclareGraphicsExtensions{.pdf}

\newcommand{\ket}[1]{|#1\rangle}
\newcommand{\bra}[1]{\langle #1 |}
\newcommand{\brkt}[2]{\langle #1 | #2 \rangle}  
\newcommand{\qsym}[2]{\bigl|\raisebox{0.33ex}{\(\underset{{{}^{#1}}}{#2}\)}\bigr\rangle}
\newcommand{\cc}[1]{{\mkern1mu\overline{\mkern-1mu #1\mkern-1mu}\mkern1mu}}
\newcommand{\mi}{i}
\newcommand{\me}{e}
\newcommand{\Id}{\openone}

\newcommand{\wb}[2]{\phi_{#1}^{#2}}
\newcommand{\kwb}[2]{\ket{\wb{#1}{#2}}} 
\newcommand{\si}[1]{\ {}_{\mathtt{[#1]}}\!\!}

\begin{document}

\title{Scheme of quantum communications based on Witting polytope}

\author{Alexander Yu.\ Vlasov}

\begin{abstract}
Currently, generalizations of quantum communication protocols from qubits to systems with higher-dimensional state spaces (qudits) typically use mutually unbiased bases (MUB). The construction with maximal number of MUB is known in 
any dimension equal to a prime power and at least two such bases exist in any dimension. However, in small 
dimensions, there also exist formally more symmetric systems of states, described by regular complex polytopes, 
which are a generalization of the idea of Platonic solids to complex spaces.

This work considers the application of a model originally proposed by R. Penrose and based on the geometry of dodecahedron and two entangled particles with spin 3/2. In a more general case, two arbitrary quantum systems with four basis states (ququarts) can be used instead. It was later shown that this system with 40 states is equivalent to the Witting configuration and is related to the four-dimensional complex polytope described by Coxeter. Presented paper describes how to use this configuration for a quantum key distribution protocol based on contextuality using some illustrative examples
with 40 ``quantum cards''.
\end{abstract}

\maketitle

\section{Introduction}\label{intro}

The model of ``Bell non-locality without probabilities'' based on geometry of dodecahedron
and two entangled particles with spin-3/2 was suggested by R. Penrose and developed
in series of publications \cite{PenDod,ZP,shadows}. So-called Majorana representation \cite{Ma32}
of arbitrary particle with spin $n/2$ as a set with $n$ points on a sphere was used
in these works.

The analysis of the model by other authors \cite{MA} produced simplified equivalent representation
of 40 quantum states used in this model without Majorana representation. Up to insignificant phase 
multipliers such transition can be described by some unitary transformation. Maybe, illustrative
relation with geometry of dodecahedron was partially lost in such representation.

However, new representation simplifies application of more general approach to the model
description and now arbitrary quantum systems with four-dimensional state space (ququart)
can be naturally used instead of only spin-3/2 particles. For such representation also
was found relation with a fairly well studied geometrical object, the 4D regular complex Witting
polytope \cite{WA}.

\smallskip

Presented work developes approach from earlier work \cite{VlDod} and discusses a 
quantum key distribution (QKD) protocol based on the models of 
Penrose dodecahedra and the Witting polytope. The Witting polytope and Witting
configuration are recollected in Section~\ref{sec:witt} along with
discussion about symmetries and two methods of states numbering.
The first one already was used in an earlier work \cite{VlDod} and more
illustrative indexing using desk with 40 ``quantum cards'' is also considered.
The {\em contextual} QKD protocol based on Witting configuration is described 
in Section~\ref{sec:QKD}. Interrelations between some classical and quantum models
are discussed in Section~\ref{sec:mod} using ``quantum cards'' for demonstration of 
some specific properties of such protocol related with idea of contextuality. 
The Section~\ref{sec:concl} summarizes some basic ideas and briefly describes
new quantum key agreement protocol derived from presented model.

\section{\label{sec:witt}Witting polytope and Witting configuration}

Regular complex polytopes were suggested by Coxeter as a generalization of regular polygons and 
polyhedra (Platonic solids) to complex spaces of various dimensions \cite{CoxReg}.
The four-dimensional complex Witting polytope has 240 vertices and can be described by expressions
\cite{CoxReg}
\begin{equation}\label{points240}
\begin{array}{lccr}
 (0,\pm\omega^\mu,\mp\omega^\nu,\pm\omega^\lambda),\quad
 &(\mp\omega^\mu,0,\pm\omega^\nu,\pm\omega^\lambda), \quad
 &(\pm\omega^\mu,\mp\omega^\nu,0,\pm\omega^\lambda),\quad
 &(\mp\omega^\mu,\mp\omega^\nu,\mp\omega^\lambda,0), \\
 (\pm\mi\omega^\lambda \sqrt{3},0,0,0),\quad
 &(0,\pm\mi\omega^\lambda \sqrt{3},0,0),\quad 
 &(0,0,\pm\mi\omega^\lambda \sqrt{3},0),\quad
 &(0,0,0,\pm\mi\omega^\lambda \sqrt{3}),
\end{array} 
\end{equation} 
where $\omega = (-1+\mi \sqrt{3})/2= \me^{2\pi\mi/3}$, $\omega^3=1$ and
$\lambda, \mu, \nu \in \{0,1,2\}$.

If, respecting principles of quantum mechanics, to do not take into account six different phase 
factors for any vector in (\ref{points240}), then we get a set of only 40 different quantum states, 
similar to the configuration described by Witting in 1887 \cite{Wit1887} in honor of which the 
polytope was named. This set includes four basic states
\begin{subequations}\label{rays40}
\begin{equation}\label{rays40b}
 (1,0,0,0),\qquad (0,1,0,0),\qquad
 (0,0,1,0),\qquad (0,0,0,1)
\end{equation}
and 36 states which can be expressed as four groups with 9 states in each
\begin{equation}
\label{rays40w}
 \frac{1}{\sqrt{3}}(0,1,-\omega^\mu,\omega^\nu),\quad
 \frac{1}{\sqrt{3}}(1,0,-\omega^\mu,-\omega^\nu), \quad
 \frac{1}{\sqrt{3}}(1,-\omega^\mu,0,\omega^\nu),\quad
 \frac{1}{\sqrt{3}}(1,\omega^\mu,\omega^\nu,0),
\end{equation}
\end{subequations}
where $\mu, \nu \in \{0,1,2\}$, $\omega  = \me^{2\pi\mi/3}$.

Regular polyhedra and polygons can be described using their group of symmetry, 
which is a discrete subgroup of the rotation group. Complex polytopes also
can be described by their group of symmetry, which is discrete subgroup of unitary group.
Rotations can be represented as compositions of reflections and similarly, discrete 
symmetries of complex polytopes can be defined via ``complex reflections'' \cite{CoxReg}.

The square of an usual reflection is identity transformation. For complex reflections
some other power $k$ (order) can correspond to identity and it can be convenient to use
Dirac notation for them
\begin{equation}
R_{\ket{\phi}} = \Id + (\me^{2\pi\mi/k}-1)\ket{\phi}\!\bra{\phi}.
\label{CxRef}
\end{equation}
In this expression $R$ depends on vector (state) $\ket{\phi}$ and normalization is supposed
$\brkt{\phi}{\phi}=1$. For $k=2$ and real vector such expression corresponds to
usual reflection.
For Witting polytope all symmetries can be expressed as compositions of
four third-order complex reflections $k=3$ ({\em triflections} \cite{Atl}),
denoted further as $R_j$, $j=1,\ldots,4$.
They can be obtained if to apply equation (\ref{CxRef}) to four vectors
\begin{equation}
 (1,0,0,0), \quad \frac{1}{\sqrt3}(1,1,1,0),\quad
 (0,0,1,0), \quad \frac{1}{\sqrt3}(0,1,-1,1).
\label{RefVec}
\end{equation}

With different compositions of $R_j$ we can obtain arbitrary symmetry of 240 vertexes
of Witting polytope. To drop common phases in quantum states the alternative set
of matrices with determinants equal to unit may be more convenient 
$R'_j = \omega^2 R_j$, $j=1,\ldots,4$.
In such a case all 40 states of Witting configuration can be constructed by
application of different compositions of $R'_j$ to state $\ket{0}$, {\em i.e.}, 
vector $(1,0,0,0)$. Such approach can be useful due to natural relation
with transformation of states by quantum gates. Symmetry group of 40 states
in Witting configuration is fairly big and includes 51840 transformations,
yet only half of them can be treated as ``truly different'' for quantum system 
due to presence of $\pm 1$ multiplier in the group.

\medskip

All 40 states are represented in Table~\ref{tab_bl} from \cite{VlDod}. 
All components are multiplied on $\sqrt{3}$ due to typographical reasons. 
It is denoted as ``block'' numbering here due to structure of algorithm
used for construction of such indexes. The algorithm for GAP system of 
computer algebra can be found in an ancillary file distributed with this work.

\begin{table*}[htbp]
\begin{center}\caption{\label{tab_bl} ``Block'' numbering}
\[
\begin{array}{|r||c|c|c|c|}
\hline
\vphantom{\Big|} n&\sqrt{3}\,\kwb0n &\sqrt{3}\,\kwb1n &\sqrt{3}\,\kwb2n &\sqrt{3}\,\kwb3n\\ \hline
0& (\sqrt{3},\ 0,\ 0,\ 0)& (0,\sqrt{3},\ 0,\ 0)&
( 0,\ 0,\sqrt{3},\ 0)& (0,\ 0,\ 0,\sqrt{3})\\ \hline
1& (0,\ 1,-1,\ 1)& (1,\ 0,-1,-1)&
(1,-1,\ 0,\ 1)& (1,\,\ 1,\,\ 1,\,\ 0)\\
2& (0,\ 1,-\omega,\ \cc\omega)&
(1,\ 0,-\omega,-\cc\omega)&
(1,-\omega,\ 0,\ \cc\omega)&
(1,\ \omega,\ \cc\omega,\ 0)\\
3& (0,\ 1,-\cc\omega,\ \omega)&
(1,\ 0,-\cc\omega,-\omega)&
(1,-\cc\omega,\ 0,\ \omega)&
(1,\ \cc\omega,\ \omega,\ 0)\\ \hline
4& (0,\ 1,-\omega,\ 1)&
(1,\ 0,-1,-\omega)&
(1,-\cc\omega,\ 0,\ \cc\omega)&
(1,\ \omega,\ 1,\ 0)\\
5& (0,\ 1,-\cc\omega,\ \cc\omega)&
(1,\ 0,-\omega,-1)&
(1,-1,\ 0,\ \omega)&
(1,\ \cc\omega,\ \cc\omega,\ 0)\\ 
6& (0,\ 1,-1,\ \omega)&
(1,\ 0,-\cc\omega,-\cc\omega)&
(1,-\omega,\ 0,\ 1)&
(1,\ 1,\ \omega,\ 0)\\ \hline
7& (0,\ 1,-\cc\omega,\ 1)&
(1,\ 0,-1,-\cc\omega)&
(1,-\omega,\ 0,\ \omega)&
(1,\ \cc\omega,\ 1,\ 0)\\
8& (0,\ 1,-1,\ \cc\omega)&
(1,\ 0,-\omega,-\omega)&
(1,-\cc\omega,\ 0,\ 1)&
(1,\ 1,\ \cc\omega,\ 0)\\
9& (0,\ 1,-\omega,\ \omega)&
(1,\ 0,-\cc\omega,-1)&
(1,-1,\ 0,\ \cc\omega)&
(1,\ \omega,\ \omega,\ 0) \\ \hline
 \end{array}
\]
\end{center}
\end{table*} 

The structure of Table~\ref{tab_bl} also reflects connection between Witting
configuration and mutually unbiased bases (MUB) in 3D. One of such basis is 
represented by vectors with last component is equal to zero from first row and 
last column in the table:
$\{\kwb00,\kwb01,\kwb02\}$, $\{\kwb31,\kwb32,\kwb33\}$, $\{\kwb34,\kwb35,\kwb36\}$,
$\{\kwb37,\kwb38,\kwb39\}$.
Such triplets of states are ``building blocks'' of the table.
All columns of this table, up to an insignificant common phase factor (also affecting
on relative placing of $\pm$ signs), can be 
obtained from each other by cyclic shifts of the vector components.

However, such an indexing is not reflect some useful symmetries of configuration.
The model looks more illustrative with a map into set with $10 \times 4 = 40$ cards of four
suits represented in Table~\ref{tab_card}.
The rule for forming this table is not based on a cyclic shift of columns, but on the 
requirement that all states in each row should be mutually orthogonal with respect to
Hermitian scalar product $\brkt\Phi\Psi$.

\begin{table*}[htbp]
\begin{center}
\caption{\label{tab_card} Indexing of states by ``quantum cards''}
(the corresponding ``block'' index is shown in square brackets)
\[
\begin{array}{|r||c|c|c|c|}
\hline
\vphantom{\Big|} n&\ \sqrt{3}\, \qsym{\spadesuit}{\mathsf{n}} &
\ \sqrt{3}\, \qsym{\heartsuit}{\mathsf{n}}&
\ \sqrt{3}\, \qsym{\diamondsuit}{\mathsf{n}} &
\ \sqrt{3}\, \qsym{\clubsuit}{\mathsf{n}}\\ \hline
\sf 1& (\sqrt{3},\ 0,\ 0,\ 0)\si0& (0,\sqrt{3},\ 0,\ \ 0)\si0&
( 0,\ 0,\sqrt{3},\ 0)\si0& (0,\ 0,\ 0,\!\sqrt{3})\si0\\ \hline
\sf 2& (0,\ 1,-1,\ \ 1\,)\si1& (1,\ 0,-1,-1)\si1&
(1,-1,\ 0,\ 1)\si1& (1,\ 1,\,\ 1,\ 0)\si1\\
\sf 3& (0,\ 1,-\omega,\ \cc\omega)\si2&
(1,\ 0,-\cc\omega,-1)\si9&
(1,-\omega,\ 0,\ 1)\si6&
(1,\ \omega,\ \cc\omega,\ 0)\si2\\
\sf 4& (0,\ 1,-\cc\omega,\ \omega)\si3&
(1,\ 0,-\omega,-1)\si5&
(1,-\cc\omega,\ 0,\ 1)\si8&
(1,\ \cc\omega,\ \omega,\ 0)\si3\\ \hline
\sf 5& (0,\ 1,-1,\ \omega)\si6&
(1,\ 0,-\omega,-\cc\omega)\si2&
(1,-\omega,\ 0,\ \cc\omega)\si2&
(1,\ \omega,\ \omega,\ 0)\si9\\ 
\sf 6& (0,\ 1,-\omega,\ 1)\si4&
(1,\ 0,-1,-\cc\omega)\si7&
(1,-\cc\omega,\ 0,\ \cc\omega)\si4&
(1,\ \cc\omega,\ 1,\ 0)\si7\\
\sf 7& (0,\ 1,-\cc\omega,\ \cc\omega)\si5&
(1,\ 0,-\cc\omega,-\cc\omega)\si6&
(1,-1,\ 0,\ \cc\omega)\si9&
(1,\ 1,\ \cc\omega,\ 0)\si8\\ \hline
\sf 8& (0,\ 1,-1,\ \cc\omega)\si8&
(1,\ 0,-\cc\omega,-\omega)\si3&
(1,-\cc\omega,\ 0,\ \omega)\si3&
(1,\ \cc\omega,\ \cc\omega,\ 0)\si5\\
\sf 9& (0,\ 1,-\omega,\ \omega)\si9&
(1,\ 0,-\omega,-\omega)\si8&
(1,-1,\ 0,\ \omega)\si5&
(1,\ 1,\ \omega,\ 0)\si6 \\ 
\sf 1\!0& (0,\ 1,-\cc\omega,\ 1)\si7&
(1,\ 0,-1,-\omega)\si4&
(1,-\omega,\ 0,\ \omega)\si7&
(1,\ \omega,\ 1,\ 0)\si4\\ \hline
 \end{array}
\]
\end{center}
\end{table*}

For example, states used in expressions (\ref{RefVec}) representing complex reflections
generating all symmetries of Witting configuration with such indexing can be represented as
$\qsym{\spadesuit}{\mathsf{1}}$, $\qsym{\clubsuit}{\mathsf{2}}$,
$\qsym{\diamondsuit}{\mathsf{1}}$, $\qsym{\spadesuit}{\mathsf{2}}$, respectively.
The advantage of the ``card'' numbering is some additional regularity in description
of 40 sets with four orthogonal states (bases) used in a model of Penrose dodecahedra
described further and represented in Tables \ref{tab_bas} and \ref{tab_bas2}.

\section{\label{sec:QKD}Quantum key distribution}

Let us consider system with two ququarts in entangled state
\begin{equation}\label{sigma}
\ket{\Sigma} = \dfrac{1}{2} \bigl(\ket{0}\ket{0}+\ket{1}\ket{1}+\ket{2}\ket{2}+\ket{3}\ket{3}\bigr).
\end{equation}

Repeated preparation of this state with distribution between two participants after measurements 
would lead to the appearance of a random sequence of pairs of identical integers with equal probability 
distributed in the range from zero to three, which could be used further for an exchange of 
encrypted messages. It should be mentioned that in Penrose model with dodecahedra, instead of this 
symmetric state, an antisymmetric one was used, but in does not affect the further consideration
in some essential way. 

The measurement with the same fixed basis $\ket{0}$, $\ldots$ , $\ket{3}$ for both participants may suffer from
eavesdropping by interception and measurement of message for one of participant with further resending 
him a state $\ket{j}$, $j=0,...,3$ depending on result of the measurement. In such a case
formally instead of the state $\ket{\Sigma}$ we have sequence of pairs $\ket{j}\ket{j}$
distributed with equal probability 1/4. To discover such a substitution the measurements in
different bases are necessary.

Let us consider quite natural extension into higher dimensional systems \cite{QDKD04} of E91 protocol initially 
suggested by A.~Ekert for pair of entangled qubits \cite{E91}. The useful property of state (\ref{sigma})
is possibility to rewrite such expression as
\begin{equation}\label{sigbas}
\ket{\Sigma} = \dfrac{1}{2} \bigl(\ket{\chi_0}\ket{\chi^*_0}
+\ket{\chi_1}\ket{\chi^*_1}+\ket{\chi_2}\ket{\chi^*_2}+\ket{\chi_3}\ket{\chi^*_3}\bigr),
\end{equation}
where $\ket{\chi_k}=\sum_{j=0}^3\chi_{kj}\ket{j}$, $j,k=0,\ldots,3$ is some orthogonal set with four states
(one of the possible alternative bases), and $\ket{\chi^*_k}$ is a ``complex conjugate'' basis,
expressed as $\ket{\chi_k^*}=\sum_{j=0}^3\cc\chi_{kj}\ket{j}$.
For particular case representing usual choice for qubits with all coefficients $\chi_{kj}$ are real 
both bases coincide. The set of coefficients $\chi_{kj}$ can be identified with matrix of transition
from standard to alternative measurement basis.

In yet another version of QKD not based on quantum entanglement and not analyzed in presented work
(prepare-and-measure QKD) MUB are often chosen as set of bases \cite{CBKG02,Kulik07}.
Similar choice looks also natural for protocols based on quantum entanglement. Really, for such a set
for any two states from different bases $|\brkt\psi\varphi|^2=1/4$ and for incorrect choice
of bases result of measurement is uniformly random and may not provide any information for eavesdropper.

For different states from Witting configuration value  $|\brkt\psi\varphi|^2$ could be equal
either to zero or to $1/3$. It complicates analysis of protocol, but for incorrect choice of
bases the result of measurement is also uniformly random, if to do not take into account
impossibility to appear as result of measurement the element corresponding to zero probability. 
Indeed, the Witting configuration could be considered as some symmetrical embedding in
4D space of four sets with 3D MUB \cite{VlDod}. It was already mentioned earlier in description
of ``block'' numbering.

Essential difference of Witting configuration from MUB or any other non-overlapping set of bases
often used in QKD is that the same state is element of {\em four different bases at once}, and 
so the total number of bases that can be composed from these states reaches 40. 
All 40 bases are represented in Table~\ref{tab_bas} using simplified ``quantum cards'' notation
and in Table~\ref{tab_bas2} with full version. Ordering and structure of tetrades
displays certain regularity unlike analogue table for ``block'' numbering represented in
earlier publication \cite{VlDod}.

Origin of such regularity may be related with rich structure of symmetries
of Witting polytope. Such symmetries can be described not only by unitary matrices
corresponding to different bases, but also with $4 \times 4$ matrices over
finite fields with either three or four elements \cite{Atl} and such structures may
appear in some way in partition into four suits and $1+3\times 3 = 10$ ranks of the cards.

\begin{table*}[htbp]
\begin{center}
\caption{\label{tab_bas}40 bases with simplified ``quantum cards'' notation}
\includegraphics[scale=1.4]{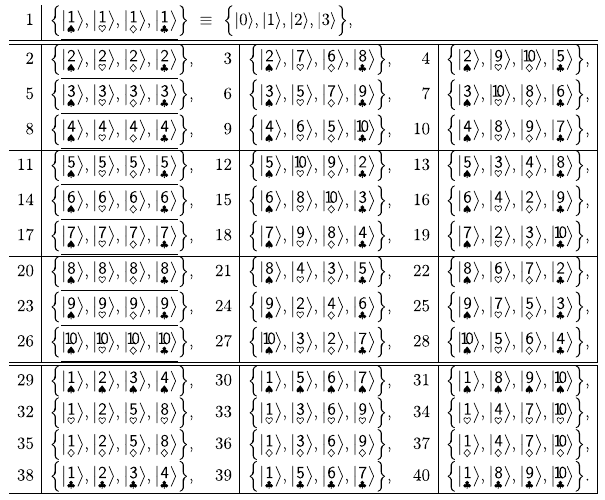} 
\end{center}
\end{table*}

\begin{table*}[htbp]
{\renewcommand\thetable{\ref*{tab_bas}$'$} 
\begin{center}
\caption{\label{tab_bas2}40 bases with full ``quantum cards'' notation}
\includegraphics[scale=1.4]{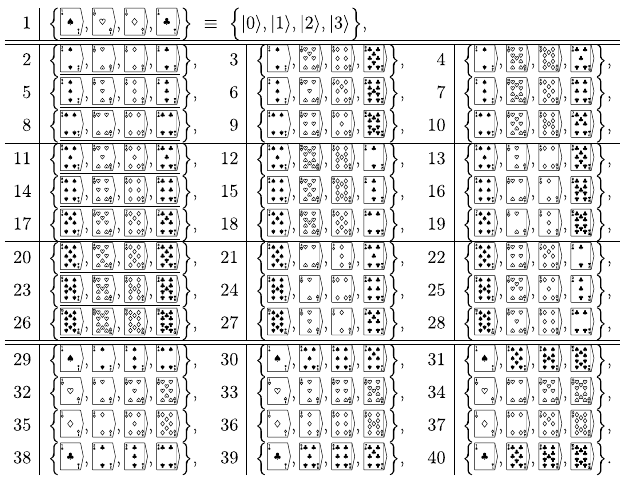} 
\end{center}
}
\addtocounter{table}{-1}
\end{table*}

\smallskip

The protocol, derived from model of Penrose dodecahedra and Witting configuration can be 
classified as {\em QKD based on contextuality}. Such approach might be understandable from
earlier works devoted to considered models \cite{PenDod,ZP,shadows,MA,WA} and discussed
further in Section~\ref{sec:mod}. 
Other examples of QKD protocols based on contextuality also exist \cite{cx11,cx17},
but suggested protocol can be also quite naturally linked with analogues and generalizations 
of E91 protocol \cite{E91,BBM92,Adv2020,Yu25}.

Let us consider in more detail the quantum communications scheme based on the Witting configuration.

Two participants (denoted further as Alice and Bob) measure entangled state $\ket{\Sigma}$
after separation into two parts and as a measurement basis $\ket{\chi_j}$, $j=0,\ldots,3$ 
presented in expression (\ref{sigbas}) the first participant uses one of 40 bases
from Table~\ref{tab_bas}/\ref{tab_bas2} and the second one uses unique defined dual basis
with conjugated states $\ket{\chi^*_j}$.

It could be suggested for convenience, that Bob uses instead of indexing from 
Table~\ref{tab_card} an alternative table with all coefficients replaced by complex conjugates.
It is sufficient to perform a rearrangement of the rows in the alternative table according to the scheme
$\sf 3{\leftrightarrow}4$, $\sf 5{\leftrightarrow}8$,
$\sf 7{\leftrightarrow}9$ and $\sf 6{\leftrightarrow}1\!0$. 
The bases for Bob also should be composed from corresponding ``conjugated'' states.
Such convention would guarantee agreement in indexing between Alice and Bob for measurements
of state $\ket{\Sigma}$ in any pairs of such ``conjugate-coordinated'' bases.

The states in each tetrade given in the Table~\ref{tab_bas}/\ref{tab_bas2} can be
associated with numbers from zero to three corresponding to the positions
within this tetrade in this table. In the first part of the table, consisting of 28 bases, 
the tetrades are indexed in the agreement with an ascending order chosen for the ``suits''
($\scriptstyle\spadesuit, \heartsuit, \diamondsuit, \clubsuit$\,),
and for the remaining 12 bases consisting of cards of the same suits --- 
in the ascending order of their numerical values.

In a certain sense, this correspondence between numbers and states is the result of a prior agreement.
The number corresponding to a given state may depend not only on it, but also on the basis.
For example, the state $\qsym{\spadesuit}{\mathsf{3}}$ would correspond to zero in three bases (5, 6, 7)
from the first part of the table, but in the fourth basis (29) it corresponds to two.
However,
$\qsym{\heartsuit}{\mathsf{3}}$ corresponds to one in all four bases (5, 13, 27, 33) in the table.

After a series of measurements in coordinated bases Alice and Bob should receive the same set of 
random numbers from zero to three, which can be used for exchange of encrypted messages.
It should be mentioned that the ``na{\'\i}ve'' version of the protocol with random selection 
of one of 40 bases leads to a rather low probability of coincidence of $2.5\%$, which can be 
increased by using more complex versions of the protocol.

One of the advantages of the contextuality-based protocols is possibility to check if the
channel indeed operates in ``quantum mode''. This means that due to malfunction of equipment
or specific attack on the protocol instead of quantum version a classical functioning
could occur. In this case, discussions about the reliability of quantum communication 
do not have any meaning. It can be treated as a failure due to ``substitution of a quantum 
protocol with a classical one''.

From a formal point of view, participants use some devices that generate identical sets of 
random numbers, and it may be necessary to ensure that their functionality is based on quantum properties.
It should be mentioned that such questions can be resolved without using contextuality.
For example, already in the E91 protocol for a similar check it was proposed to use additional 
measurements testing Bell's inequalities \cite{E91}.
However, the protocol considered here based on the Penrose model ``without probabilities''
may have some advantages.

\section{\label{sec:mod}Quantum and classical models}

To analyze the possible problem with ``substitution of a quantum protocol with a classical one'' 
that could violate the security of information transfer, let us consider the classical 
versions of the model associated with suggested protocol.
For example, Alice and Bob have two equal desks of 40 cards and both can take one of the 40 tetrades 
listed in Table~\ref{tab_bas}/\ref{tab_bas2}.
After this, as a result of some manipulations, one of these four cards is selected and 
if the protocol were to function correctly, such card should {\em always} match for both participants 
(if they selected the same four cards from their decks).

If Alice and Bob make the choice of cards independently and may not communicate to coordinate their actions 
in any way, but in any such choice the card must be the same, then we can conclude that the cards 
are marked in advance, which allow the desired card to be selected. To reproduce correct behavior 
of the quantum model, it is necessary that the distribution of marks over the 40 cards be made 
in such a way that in any tetrades from the Table~\ref{tab_bas}/\ref{tab_bas2} there is one and only 
one marked card.

However, such distribution of marks is not possible. In the Penrose model, the demonstration 
of this contradiction relied on geometric constructions, but it is sufficient for verification 
simply to consider quadruples of states, similar to that given in Table~\ref{tab_bas}/\ref{tab_bas2}.
Let us show using this example how to prove the impossibility of choosing the required distribution of marks.

First, let us consider only the ten tetrades underlined in the Table~\ref{tab_bas}/\ref{tab_bas2} 
(1, 2, 5, 8, 11, 14, 17, 20, 23, 26) and 
consisting of four cards of the same rank in different suits. Obviously, each of the 40 cards appears in these 
ten tetrades once and only once, and if instead of the 40 bases (tetrades) only these ten were used, 
then the placement of marks would become quite trivial. It would be necessary to simply place at random 
a mark on only one of the cards in each of these ten tetrades.

However, this approach to marking can lead to incorrect distribution of marks 
in the remaining 30 tetrades. For example, suppose that in all ten tetrades the same suit was chosen, 
for example, spades. In this case, choosing any of the 28 tetrades from the first part of the table
(i.e., in $70\%$ cases) leads to the correct result, but in the remaining 12 tetrades, all four cards 
are either marked or have no marks.

Next, the selected ten tetrades of cards can be used as a core for description of $4^{10}=1048576$ 
possible distributions of marks because any correct distribution should have one and only one mark in
any tetrade. One can see how marks are arranged in the remaining 30 tetrades. A brute 
force enumeration confirms that there will always be tetrades of cards for which the distribution of 
marks will not be correct. 
The calculation also shows that with a random selection, the probability of getting a correct tetrade
of cards with one mark is about $57\%$, which is even less than in the example above.

The maximum possible number of correct tetrades is 34 (i.e., $85\%$ cases). An example is the arrangement of labels corresponding to the states
\[\left\{
 \qsym\spadesuit{\sf 1}, \qsym\spadesuit{\sf 2}, 
 \qsym\heartsuit{\sf 3}, \qsym\heartsuit{\sf 4}, \qsym\heartsuit{\sf 5},
 \qsym\clubsuit{\sf 6}, 
 \qsym\diamondsuit{\sf 7}, \qsym\diamondsuit{\sf 8}, \qsym\diamondsuit{\sf 9}, 
 \qsym\clubsuit{\sf 1\!0}\right\}\!,
\]
in which in 34 bases only one state is marked as required,
but in three bases there will be two marks at once and in the other three --- no one.
In Table~\ref{tab_bas}/\ref{tab_bas2} these are bases with numbers (6, 7, 29) and (15, 25, 38), 
respectively. There are only 720 such sets and their percentage among all possible
$4^{10}$ variants is less than $0.07\%$, and the percentage of combinations for which
the number of correct tetrades exceeds $70\%$ is
less than $4\%$. So when using this type of classical
models, the violation of the necessary requirements will appear after a not very large
number of tests with different combinations of cards.

\smallskip

However, correct results can be reproduced using contextual classical models. For example,
instead of one desk of 40 cards given to each participant, four such desks can be used, 
from which all 40 necessary combinations given in Table~\ref{tab_bas}/\ref{tab_bas2}
can be composed from very beginning, and in each of them, the same cards in the desks of
both participants must be marked. In this case, each participant does not choose one card, 
but immediately takes one of the 40 tetrades of cards and finds the one marked in advance.

This model is even closer to the usual measurement procedure used in the formalism of 
quantum circuits and their simulators, in which intermediate measurements 
are usually not allowed, which leads to some difficulty with the implementation of the idea 
used by Penrose. But the implementation of just described model is formally contextual,
since the same card can be either marked or not, depending on other three cards included
in the same tetrade. This situation can be avoided if we would require the conformance
to a certain measurement procedure, similar to that used in the model considered by 
Penrose \cite{PenDod,ZP,shadows}.

The implementation of this procedure in a modified version of this model
based on the Witting configuration has already been considered earlier \cite{VlDod}.
For this, at the first step it is necessary to select one of 40 states and make a measurement,
the result of which can be interpreted as an answer to the question of whether
the system is in this state or not and only after that select one of the four bases
to which this state belongs and complete the measurement procedure, having received
as a result the final information about the state of the system.

In the Penrose model, such a step-by-step measurement scheme is proposed to be implemented 
in a way similar with {\em path-encoding} in quantum communications. However, 
the corresponding measurement procedure is not very natural in the standard approach using
quantum circuits, so two auxiliary qubits may be required to implement the first step of
the measurement \cite{VlDod}.

In this case, in order to use the standard measurement procedure adopted in the theory
of quantum circuits, one can use the idea of a delayed query, in which instead of measuring
the state interacts with an auxiliary qubit.
Each such qubit corresponds to a ``non-destructive'' (delayed) query,
whether the system (ququart) is in a certain predetermined state $\ket{\psi}$ and
the unitary operator performing the necessary interaction can be expressed as
\begin{equation}\label{measgate}
 T_\psi = \ket{\psi}\bra{\psi}\otimes X + \bigl(\Id - \ket{\psi}\bra{\psi}\bigr) \otimes \Id.
\end{equation}
If to encode a ququart as a set of two qubits, then the simplest example of such an
operator is the Toffoli gate, corresponding to $\ket{\psi}$ is equal to
$\ket{1}\ket{1}$ for two qubits or $\ket{3}$ for ququart.

The approach with auxiliary qubits allows for easier modeling of this protocol
on simulators. It may seem, in the ``na{\'\i}ve'' version of the protocol in such a two-step approach,
the already mentioned problem with low efficiency manifests itself even more strongly.
The probability that both participants will choose the same state is $1/40$,
and after that, one of four bases is selected, bringing the probability of coincidence
to $1/160$.
However, the probability that after a two-step procedure both participants to choose the same 
basis is $1/40$, and even though in $3/4$ cases they chose different states at the first step, the coordinated 
basis always allows them to use the result of the measurements to generate the same key.

\section{\label{sec:concl}Conclusion and discussion}

In this paper, a quantum entanglement-based key distribution protocol using contextuality was considered.
The original idea of the protocol is based on the model proposed by Penrose and based on the geometry of 
the dodecahedron. The Witting configuration simplifies the expressions for quantum states, but loses 
some clarity associated with the direct connection with the geometry of dodecahedra.

To partially correct this shortcoming, a state numbering scheme was chosen that uses a desk with 40 cards of 
four suits and ranks from one (``ace'') to ten.
As can be seen from Table~\ref{tab_bas}/\ref{tab_bas2}, the tetrades of cards corresponding to the 40 bases have a 
certain regularity that is absent from the alternative indexing previously used in \cite{VlDod}.

The work did not address issues related to the practical implementation of the protocol. For example, a 
separate topic deserves an analysis of which quantum systems can be used as ququarts.

The low efficiency of the protocol, inversely proportional to the number of bases searched by the participants 
during the exchange, was already mentioned. It should be noted that the measurement using all 40 sets is 
only necessary to check the conditions related with contextuality. In addition, even in this case, the 
minimum required number of states in the original model can be reduced to 28 \cite{ZP}.

In addition, in a simplified version of the protocol, maximum efficiency can be achieved by assuming that, 
although both the choice of states and the choice of bases are random, they can be agreed upon by both 
participants. For example, they can both use the same (pseudo)random number generator. Although it is difficult 
to imagine how such a version could be used for secret communications, it could be useful for experiments related to fundamental questions in quantum theory. An intermediate case, where the choices made by both participants are not necessarily the same, but are correlated in some way to increase the probability of a
coincidence, could also be useful.

\smallskip

{\em Quantum key-agreement protocol.}
Another possibility to greatly improve efficiency is related to the modification of protocol briefly described below. 
As already noted, in the original two-step version of the protocol, participants first use a check to see if the 
system is in one of 40 states.

Let us suppose that the participants independently selected two states for this purpose
and they have the opportunity to postpone the second measurement and during this time exchange 
information about which states they selected. In this case, the probability that both states 
may belong to the same basis between 40 is $13/40$ ({\em i.e}, $32.5\%$, where $2.5\%$ is the 
probability to choose the same state and $30\%$ is the probability for two different states 
falling into the same basis). In this case, the participants can use this basis in the second 
step to obtain consistent measurement results.

The approach just described is more similar to {\em a key-agreement protocol}, although in classical 
cryptography this corresponds to different principles. From a technical point of view, the implementation 
of such a quantum key agreement protocol may be complicated, since it requires a delay between two
stages of the quantum protocol necessary for the exchange of classical messages carrying information about the 
selected states. It should be mentioned that interception of only these messages does not lead to information leakage, 
since it is unknown what specific measurement results will be obtained after this exchange and
the selection of an agreed basis.

\begin{acknowledgments}
Author expresses gratitude to organizers and sponsors of XII Klyshko seminar for possibility of participation 
and presentation of invited talk (26 October 2024), on which this paper was based.
\end{acknowledgments}


\begin{thebibliography}{99}
\bibitem{PenDod} {R. Penrose}, 
``On Bell non-locality without probabilities: some curious geometry,''
 Preprint, Oxford, 1992; reprinted in: 
\textit{J.~Ellis, D. Amati}, Quantum Reflections, 1--27, CUP, 2000. 
\bibitem{ZP} {J. Zimba, R. Penrose}, 
``On Bell non-locality without probabilities: more curious geometry,''
\href{https://doi.org/10.1016/0039-3681(93)90061-N}{Stud. Hist. Phil. Sci. {\bf 24}, 697--720 (1993)}.
\bibitem{shadows} {R. Penrose}, Shadows of the Mind, OUP, 1994.
\bibitem{Ma32} {E. Majorana},
``Atomi orientati in campo magnetico variabile,''
\href{https://doi.org/10.1007/BF02960953}{Nuovo Cimento {\bf 9}, 43--50 (1932)}.  
\bibitem{MA} {J. Massad, P. K. Aravind}, 
``The Penrose dodecahedron revisited,''
\href{https://doi.org/10.1119/1.19336}{Am. J. Phys. {\bf 67}, 631--638 (1999)}.
\bibitem{WA} {M. Waegell, P. K. Aravind},
``The Penrose dodecahedron and the Witting  polytope are identical in $\mathbb{CP}^3$,'' 
\href{https://doi.org/10.1016/j.physleta.2017.03.039}{Phys. Lett. A {\bf 381}, 1853--1857 (2017)},
 Erratum: \href{https://doi.org/10.1016/j.physleta.2020.126331}{Phys. Lett. A {\bf 384}, 126331 (2020)}. 
\bibitem{VlDod}{A. Y. Vlasov},
``Penrose dodecahedron, Witting configuration and quantum entanglement,''
\href{https://doi.org/10.48550/arXiv.2208.13644}{Preprint arXiv:2208.13644 (2022)};
\href{https://doi.org/10.12743/quanta.v13i1.276}{Quanta {\bf 13}, 38--46 (2024)}.
\bibitem{CoxReg} {H. S. M. Coxeter},  Regular Complex Polytopes, CUP, 1991.
\bibitem{Wit1887}{A. Witting}, 
``Ueber Jacobi'sche  Functionen  k$^{ter}$  Ordmmg  zweier Variabler,''
\href{https://eudml.org/doc/157278}{Math. Ann. {\bf 29}, 157--170 (1887)}. 
\bibitem{Atl} J. H. Conway, R. T. Curtis, S. P. Norton, R.~A.~Parker, and R.~A.~Wilson,
ATLAS of Finite Groups, 26--27, OUP, 1985. 
\bibitem{E91}A. Ekert,
``Quantum cryptography based on Bell's theorem,''
\href{https://doi.org/10.1103/PhysRevLett.67.661}{Phys. Rev. Lett. {\bf 67}, 661--663 (1991)}.
\bibitem{QDKD04}T. Durt, D. Kaszlikowski, J.-L. Chen, L.C. Kwek,  
``Security of quantum key distributions with entangled qudits,''
\href{https://doi.org/10.1103/PhysRevA.69.032313}{Phys. Rev. A {\bf 69}, 032313 (2004)}.
\bibitem{CBKG02}N. J. Cerf, M. Bourennane, A. Karlsson, and N. Gisin, 
``Security of quantum key distribution using d-level systems,''
\href{https://doi.org/10.1103/PhysRevLett.88.127902}{Phys. Rev. Lett. {\bf 88}, 127902 (2002)}.
\bibitem{Kulik07} 
S. P. Kulik, A. P. Shurupov,
``On quantum key distribution using ququarts,''
\href{https://doi.org/10.1134/S106377610705007X}{J. Exp. Theor. Phys. {\bf 104}, 736--742 (2007)}.
\bibitem{cx11}A. Cabello, V. D'Ambrosio, E. Nagali, F. Sciarrino,
``Hybrid ququart-encoded quantum cryptography protected by Kochen-Specker contextuality,''
\href{https://doi.org/10.1103/PhysRevA.84.030302}{Phys. Rev. A {\bf 84}, 030302(R) (2011)}.
\bibitem{cx17}J. Singh, K. Bharti, Arvind,
``Quantum key distribution protocol based on contextuality monogamy,''
\href{https://doi.org/10.1103/PhysRevA.95.062333}{Phys. Rev. A {\bf 95}, 062333 (2017)}.
\bibitem{BBM92} C. H. Bennett, G. Brassard, and N. D. Mermin, 
``Quantum cryptography without Bell’s theorem,'' 
\href{https://journals.aps.org/prl/abstract/10.1103/PhysRevLett.68.557}%
{Phys. Rev. Lett. {\bf 68}, 557--559 (1992)}.
\bibitem{Adv2020}M. Erhard, M. Krenn, A. Zeilinger,
``Advances in high dimensional quantum entanglement,''
\href{https://doi.org/10.1038/s42254-020-0193-5}{Nature Rev. Phys. {\bf 2}, 365--381 (2020)}.
\bibitem{Yu25} H. Yu, S. Sciara, M. Chemnitz, {\em et al.}, 
``Quantum key distribution implemented with d-level time-bin entangled photons,'' 
\href{https://doi.org/10.1038/s41467-024-55345-0}{Nat. Commun. {\bf 16}, 171 (2025)}. 
\end{thebibliography}
\end{document}